\documentclass[reprint, aps,pra,groupeaddress,longbibliography]{revtex4-2}

\usepackage{graphicx}	% Include figure files
\usepackage{dcolumn}	% Align table columns on decimal point
\usepackage{bm}		% bold math
\usepackage{SIunits}	% Pour utiliser les unités physique
\usepackage{xcolor}
\usepackage{booktabs}
\usepackage{float}
\usepackage{ulem}
\usepackage{hyperref}
\definecolor{r}{rgb}{0.86,0.08,0.23}

\definecolor{blue_n}{rgb}{0.,0.3,0.5}

\hypersetup{
backref=true, 
pagebackref=true,
hyperindex=true, 
colorlinks=true, 
breaklinks=true, 
citecolor = blue_n, 
urlcolor= blue_n,
linkcolor= black, 
pdftitle={Single G centers in silicon fabricated by co-implantation with carbon and proton}, 
pdfauthor={Yoann~Baron,  Alrik~Durand, Tobias~Herzig, Mario~Khoury, S\'ebastien~Pezzagna, Jan~Meijer, Isabelle~Robert-Philip, Marco~Abbarchi, Jean-Michel Hartmann, Shay Reboh, Jean-Michel~G\'erard, Vincent~Jacques, Guillaume~Cassabois and Ana\"is~Dr\'eau } 
}

\begin{document}

\title{Single G centers in silicon fabricated by co-implantation with carbon and proton}

\author{Yoann~Baron$^{1}$}	 
\thanks{These authors contributed equally to this work.}	
\author{Alrik~Durand$^{1}$}
\thanks{These authors contributed equally to this work.}
\author{Tobias~Herzig$^2$}
\author{Mario~Khoury$^3$}			
\author{S\'ebastien~Pezzagna$^2$}		
\author{Jan~Meijer$^2$}		
\author{Isabelle~Robert-Philip$^1$}	
\author{Marco~Abbarchi$^3$}
\author{Jean-Michel Hartmann$^4$}
\author{Shay Reboh$^4$}		
\author{Jean-Michel~G\'erard$^5$}	
\author{Vincent~Jacques$^1$}		
\author{Guillaume~Cassabois$^1$}		
\author{Ana\"is~Dr\'eau$^1$}		
\email{anais.dreau@umontpellier.fr}	

\affiliation{$^1$Laboratoire Charles Coulomb, Universit\'e de Montpellier and CNRS, 34095 Montpellier, France} 
\affiliation{$^2$Division of Applied Quantum Systems, Felix Bloch Institute for Solid State Physics, University Leipzig, Linn\'estra\ss e 5, 04103 Leipzig, Germany}
\affiliation{$^3$CNRS, Aix-Marseille Universit\'e, Centrale Marseille, IM2NP, UMR 7334, Campus de St. J\'er\^ome, 13397 Marseille, France }
\affiliation{$^4$Univ. Grenoble Alpes and CEA, LETI,  F-38000 Grenoble, France} 
\affiliation{$^5$Univ. Grenoble Alpes, CEA, Grenoble INP, IRIG-PHELIQS, F-38000 Grenoble, France}

\begin{abstract}
We report the fabrication of G centers in silicon with an areal density compatible with single photon emission at optical telecommunication wavelengths. Our sample is made from a silicon-on-insulator wafer which is locally implanted with carbon ions and protons at various fluences. Decreasing the implantation fluences enables to gradually switch from large ensembles to isolated single defects, reaching areal densities of G centers down to $\sim$0.2 $\mu$m$^{-2}$. Single defect creation is demonstrated by photon antibunching in intensity-correlation experiments, thus establishing our approach as a reproducible procedure for generating single artificial atoms in silicon for quantum technologies.
\end{abstract}
\maketitle

% INTRO
\def\thefootnote{*}\footnotetext{These authors contributed equally to this work}\def\thefootnote{\arabic{footnote}}
%%%%%%%%%%%%%%%%%%%%%%%%%%%%%%%%%%%%%%%%%%%%%%%%%%%%%%%%%%%%%%%%%%%%%%%%%%
Silicon is the major semiconductor of the information society. It is at the heart of the devices in microelectronics and computer technology, and as such the most desired platform for the development of next generation applications in quantum technologies. On the one hand, individual dopants \cite{he_two-qubit_2019, morello_donor_2020} and gate-defined quantum dots \cite{watson_programmable_2018, zwerver_qubits_2022} have already emerged for implementing electrical qubits in silicon. 
On the other hand, the literature is still very sparse on silicon-based quantum devices harnessing individual optically-active qubits for quantum communications and quantum integrated photonics \cite{simmons_single_2020,zhang_material_2020}.
Recent studies demonstrating the coherent control of fluorescent spin-defects in silicon~\cite{bergeron_silicon-integrated_2020, kurkjian_optical_2021} and the single-photon emission from several families of isolated single near-infrared color centers in silicon~\cite{redjem_single_2020,durand_broad_2021,hollenbach_engineering_2020,baron_detection_2021} could change the game. 
However, no procedure that permits to create reproducibly individual optically-active artificial atoms with various densities in silicon has been established yet. 
This capability is essential not only for optical experiments at the single-defect level, but also in view of the deterministic integration of an individual color-center in an optical photonic microstructure to build quantum photonic devices.  

In this Letter, we report on the fabrication at single-defect scale of the G center in silicon, whose emission spectrum is characterized by a zero-phonon line (ZPL) at 1279 nm matching the O-band of optical telecommunication wavelengths \cite{davies_optical_1989}. 
The microscopic structure of this defect consists of two substitutional carbon atoms connected by an interstitial silicon atom~\cite{udvarhelyi_identification_2021}. 
To create this color center, a SOI wafer is locally implanted with carbon ions and irradiated with protons at various fluences in a cross-implantation scheme. Decreasing the implantation fluences enables to gradually switch from dense to dilute arrays, reaching areal densities of G centers down to $\sim$0.2 $\mu$m$^{-2}$, while keeping the same emission spectra whatever the density. Single defect creation is demonstrated by photon antibunching in intensity-correlation experiments, thus establishing our approach as a reproducible procedure for generating single artificial atoms in silicon for quantum technologies.
%%%%%%%%%%%%%%%%%%%%%%%%%%%%%%%%%%%%
\begin{figure}[h!]
  \includegraphics[width=1\columnwidth]{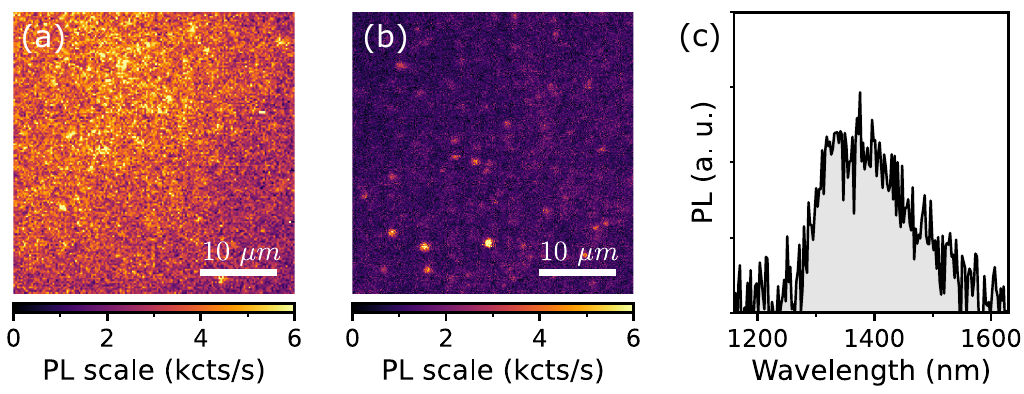}
  \caption{PL map of (a) the virgin sample (\# 2) and (b) the sample after a flash annealing of 20 s at 1000\textcelsius\, under N$_2$ atmosphere (\# 3). 
  (c) Typical PL spectrum of the hotspots in (a) and (b). Data are recorded at 30 K under excitation at 532 nm with an optical power of 10 $\mu$W.}
  \label{fig1}
\end{figure}
%%%%%%%%%%%%%%%%%%%%%%%%%%%%%%%%%%%%
%%%%%%%%%%%%%%%%%%%%%%%%%%%%%%%%%%%%
\begin{figure}[h!]
  \includegraphics[width=0.9\columnwidth]{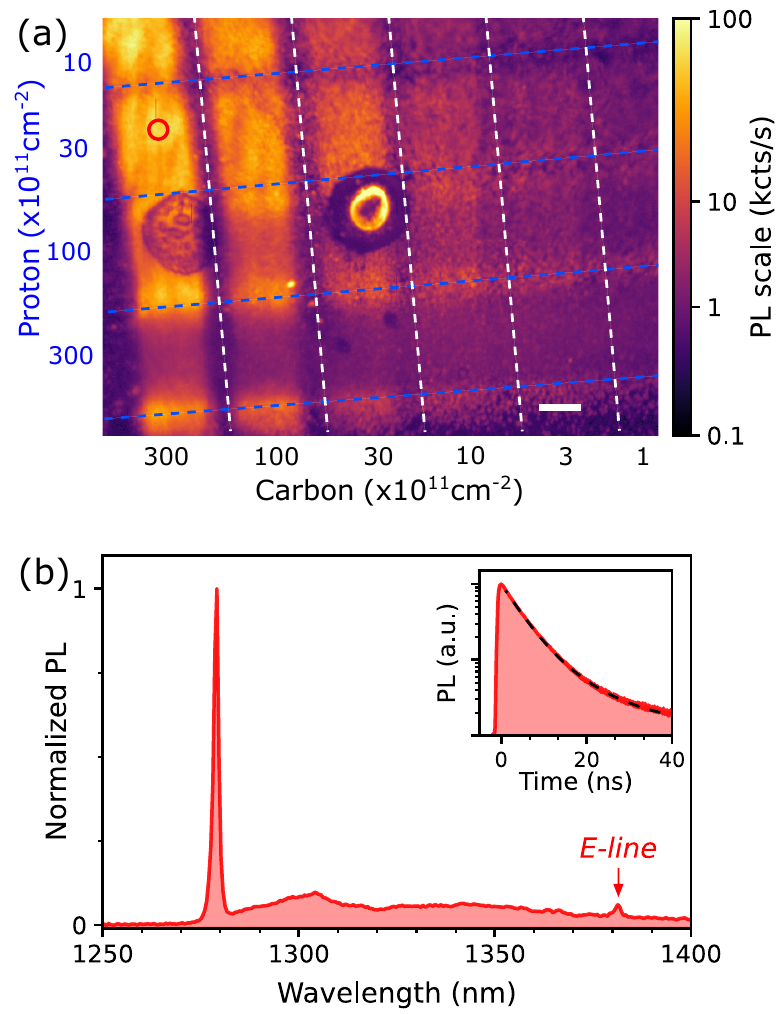}
  \caption{(a) PL map of the SOI sample (\# 1) cross-implanted with carbon and proton in the region of high irradiation fluences. 
    The vertical white (horizontal blue) dashed lines are guides for the eyes splitting the stripes with different carbon (proton) fluences, whose values are indicated in the bottom (left) axis. 
  The white bar scales for 10 $\mu$m. 
  The two 20 $\mu$m large spots visible in the left middle of the scan are due to local sample imperfections. 
  (b) PL spectrum detected for an ensemble of G centers at the position indicated by the red circle in (a).
These data are recorded at 30 K under excitation at 532nm with an optical power of 10 $\mu$W. 
  Inset: Time-resolved PL decay recorded with a 150 ps pulsed laser at 532 nm. 
  The black dashed line represents data fitting with a bi-exponential function (see main text for details). }
  \label{fig2}
\end{figure}
%%%%%%%%%%%%%%%%%%%%%%%%%%%%%%%%%%%%
%%%%%%%%%%%%%%%%%%%%%%%%%%%%%%%%%%%%

%%%%%%%%%%%%%%%%%%%%%%%%%%%%%%%%%%%%

The investigated wafer consists of a $^{28}$Si epilayer grown on a commercial SOI wafer. Although isotopic purification plays no role in the present paper, the use of $^{28}$Si ensures a nuclear spin-free host matrix \cite{saeedi_room-temperature_2013} for the optically-active defects and also a reduced inhomogeneous broadening \cite{chartrand_highly_2018} for future studies. The top Si-layer of the commercial SOI is thinned down to 4 nm by thermal oxidation, followed by wet hydrofluoric acid chemical etching. 
The growth of $^{28}$Si by chemical vapor deposition is described in Ref.\cite{mazzocchi_99992_2019}. 
The resulting stack is made of a 56 nm thick layer of $^{28}$Si and a 4 nm thick layer of natural Si, that are separated from the substrate by a 145 nm thick layer made of natural silicon oxide.
This wafer is then cut into pieces to produce the samples studied here. 
To create G centers, a first sample (\# 1) is implanted with carbon ions and then irradiated with protons \cite{beaufils_optical_2018, berhanuddin_co-implantation_2012}. 
In order to heal the silicon lattice from the implantation damage, this sample is subjected to a flash annealing at 1000\textcelsius\,during 20 s under N$_2$ atmosphere in between the two implantation processes \cite{beaufils_optical_2018, berhanuddin_co-implantation_2012}. 
To serve as a reference, a second sample (\# 2) is kept in its virgin state, i.e. without any additional step following the $^{28}$Si epilayer growth. 
To analyse the influence of the annealing step, another virgin sample (\# 3) undergoes only the flash annealing without any implantation. 

The samples are characterized by spatially-resolved photoluminescence (PL) spectroscopy in a scanning confocal microscope built in a closed-cycle He cryostat, as described in Ref.\cite{redjem_single_2020}.
PL detection is performed with superconducting single-photon detectors (SingleQuantum) after a 1050 nm long-pass filter. 
Without further indications, the data presented here are recorded at 30 K under CW 532 nm excitation at 10 $\mu$W.

Before investigating the implanted SOI sample, we first analyse the virgin reference SOI sample and the influence of the flash annealing step. 
A PL map of the virgin sample \#2 is displayed in Fig.\ref{fig1}(a). 
It reveals a spatially uniform background with few hotspots of slightly higher intensity. 
Their typical PL spectrum is shown in Fig.\ref{fig1}(c): it features a broad PL emission centered at $\sim$1350 nm and no ZPL. 
Such an emission spectrum differs from the one of G centers [see Fig.\ref{fig2}(b)] \cite{beaufils_optical_2018}. 
The presence of native G centers in the pristine SOI wafer studied here can thus been ruled out. 
The impact of the flash annealing at 1000\textcelsius\,during 20 s under N$_2$ atmosphere is seen in the PL map of sample \# 3 in Fig.\ref{fig1}(b). 
The intensity of the spatially uniform PL background is reduced by a factor $\sim$2 so that the native hotspots are better resolved. 
Still, neither the intensity nor the spectrum of their PL signal changes due to the flash annealing step. 
From this second control experiment, we conclude that flash annealing alone does not contribute to the formation of G centers in silicon. 
Conversely, all the G centers studied below are the products of the implantation by both carbon ions and protons, especially in the low-density regions of the samples where single G defects are detected.

\begin{figure}[h!]
  \includegraphics[width=0.9\columnwidth]{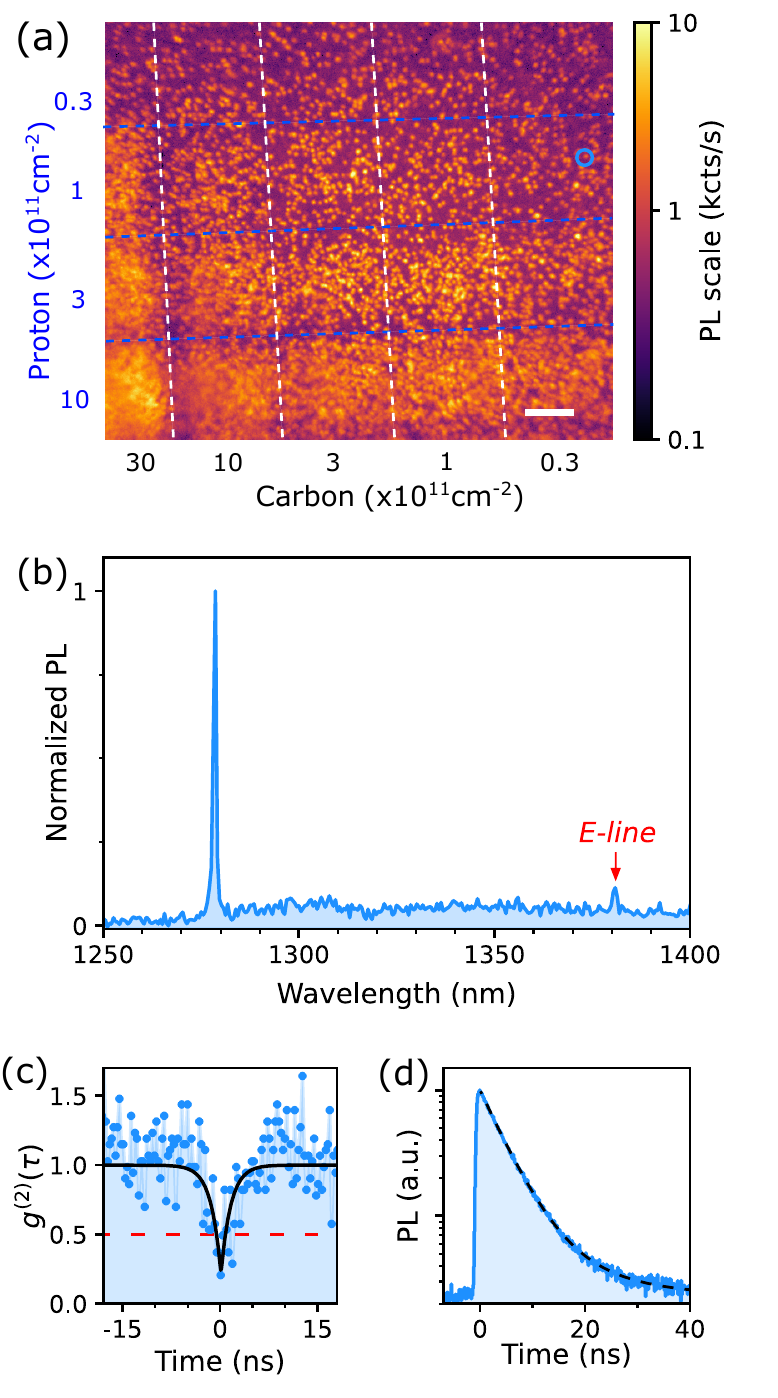}
  \caption{(a) PL map of the SOI sample \# 1 in the region of low irradiation fluences. 
   The vertical white (horizontal blue) dashed lines are guides for the eyes splitting the stripes with different carbon (proton) fluences, whose value is indicated in the bottom (left) axis. 
  The white bar scales for 10 $\mu$m. 
  (b) PL spectrum detected for an isolated G center indicated by the blue circle in (a). 
  (c) Corresponding intensity-correlation function $g^{(2)}(\tau)$, without any correction from background or detector noise counts. The photon antibunching with $g^{(2)}(0)<0.5$ evidences the emission of single photons.
  These data are recorded at 30 K under excitation at 532 nm with an optical power of 10 $\mu$W. 
  (d) Time-resolved PL trace for this single G center acquired with a 150 ps 532 nm pulsed laser.}
  \label{fig3}
\end{figure}

Dense ensembles of G centers can be created by implanting crystalline silicon with carbon ions and protons \cite{beaufils_optical_2018, berhanuddin_co-implantation_2012}. 
Our approach consists here in keeping this dual implantation while tuning the areal density of the centers by varying the implantation doses.
Localized ion implantation is performed through a 30 $\mu$m thick mica mask with a {20$\times$200~$\mu$m$^2$} aperture on sample \# 1. 
A 7$\times$7 implantation grid pattern is performed by superimposing 7 vertical carbon-implanted stripes with 7 horizontal proton-irradiated stripes separated by $\simeq$10~$\mu$m. 
 This configuration allows to probe 49 combinations of carbon and proton doses for generating G centers. The fluences vary from 0.3 to 300$\times$10$^{11}$ cm$^{-2}$ for both carbon and proton, which are implanted at energies of 8 and 6 keV, respectively.
At these energies, the carbon ions stop in the upper silicon layer whereas the protons cross it while producing interstitials and stop in the oxide layer below.

Fig.\ref{fig2}(a) shows the PL intensity map obtained from the most heavily implanted part of our SOI sample. 
The carbon and proton fluences, indicated along the bottom and left axes, vary from 1 to 300$\times$10$^{11}$cm$^{-2}$.
We observe that the sample luminescence varies significantly from one implantation square to another. 
 While the PL signal intensity increases monotonously with the carbon dose [right to left], there is a maximum at 30$\times$10$^{11}$cm$^{-2}$ when raising the proton concentration [top to bottom]. Above this fluence, the irradiated stripes acquire an inverted contrast in PL mapping with a dark center and bright horizontal edges. 
 This reveals the detrimental effect of the proton irradiation at high fluences to create G centers, as previously observed for silicon wafers highly implanted with carbon ions~\cite{berhanuddin_co-implantation_2012}. 
 
A PL spectrum recorded in the area with the highest PL intensity is displayed in Fig.\ref{fig2}(b). 
 The sharp ZPL at 1279 nm and the broad phonon-sideband at longer wavelengths are characteristic of the G center in silicon, as previously reported in ensemble measurements \cite{beaufils_optical_2018}. 
 Of particular importance is the so-called E-line at 1380 nm, which corresponds to a local vibration mode of the G center and is thus a key fingerprint of this point defect \cite{davies_optical_1989}. 
 These results confirm the efficiency of this dual carbon and proton implantation method to produce G-center ensembles in a SOI wafer. 
 Time-resolved PL experiments under 150 ps pulsed laser excitation at 532 nm [Fig. \ref{fig2}(b), inset] show that the luminescence decays first with a short constant time of 4.5 $\pm$ 0.1 ns, close to the mono-exponential decay time of $\simeq$5.9 ns reported on dense ensembles of G centers in a 220 nm thick SOI wafer \cite{beaufils_optical_2018}. 
 Nevertheless, we note that the recombination dynamics here involves also a longer decay time of 11.6 $\pm$ 0.2 ns. 
 As discussed in the following, this longer lifetime is presumably associated with other defects emitting in the spectral range of the G center.

Upon decreasing the carbon and proton fluences, the PL signal intensity decreases, as expected for more and more dilute ensembles of fluorescent defects. 
Fig.\ref{fig3}(a) shows a map of the PL signal in the low irradiation region of our cross-implanted SOI. 
In these areas, the PL signal intensity is no longer spatially uniform featuring separated bright spots whose relative distance increases while decreasing carbon and proton fluences.
The PL spectrum of such an isolated spot in the area implanted with carbon and proton respectively at $0.3\times10^{11}$ and $1\times10^{11}$ cm$^{-2}$ is displayed on Fig. \ref{fig3}(c).
It is identical to the one previously measured on the ensemble of G centers [Fig.\ref{fig2}(b)].  
The photon antibunching observed in intensity-correlation experiments with $g^{(2)}(0)<0.5$ attests that it is a single emitter [Fig.\ref{fig3}(b)] \cite{beveratos_room_2002}, thus demonstrating the detection of a single G center in silicon. 
The estimated areal density of fluorescent defects is roughly 0.2 per $\mu$m$^2$. 
We highlight the difference with our previous measurements on single defects in silicon \cite{redjem_single_2020,durand_broad_2021} where the PL spectrum was never exactly identical to the one of G center-ensembles \cite{beaufils_optical_2018}.
Besides an average shift of the ZPL of 9 nm towards the lowest wavelengths, the E-line was notably missing for the single defects reported in Ref.\cite{redjem_single_2020,durand_broad_2021}, suggesting a perturbed conformation with respect to the G center structure. 
In these former works, the SOI wafer was full scale implanted with carbon at a fluence of 500$\times$10$^{11}$cm$^{-2}$, a high value which appears unfavorable to the formation of genuine G centers. In contrast, the PL spectrum is here exactly the same for ensembles and isolated spots, as seen in Fig.\ref{fig2}(b) and Fig.\ref{fig3}(b). 
We have thus achieved the regime of single G centers in the dilute regions of the cross-implanted SOI, thereby establishing the effectiveness of this approach to produce large numbers of isolated G centers in silicon devices. 

To probe the recombination dynamics, time-resolved PL measurements under pulsed-laser excitation are reproduced on single G centers. 
As displayed on Fig.\ref{fig3}(d), the PL signal decay is mainly monoexponential with more than 80\% of the photons linked to a timescale of 4.5 $\pm$ 0.1 ns, identical to the short lifetime previously measured on the G-center ensemble [Fig.\ref{fig2}(b), inset]. 
The remaining of the photons are associated with a decay time of 12 $\pm$ 3 ns. 
Unlike the 4.5-ns time, this long decay time is still observed by repeating the experiment a few microns away from the PL isolated hotspots, suggesting it results from sample PL background. 
The slight difference in the G-center excited-state lifetime measured here and in a former sample ($\simeq$ 5.9 ns \cite{beaufils_optical_2018}) could be linked to a change of the thickness of the SOI stack layers \cite{ho_spontaneous_1993}. 
Complementary experiments will be required to further elucidate the recombination dynamics of the G center in silicon and identify its intrinsic channels.

In summary, we report the generation of single G centers in silicon by co-implantation with carbon ions and protons. Control experiments in the SOI wafer either pristine or after only flash annealing at 1000\textcelsius\,rule out the existence of native centers or their creation by the flash annealing step. G centers are solely generated by carbon and proton implantation. Decreasing the implantation fluences leads to the fabrication of dilute arrays of centers down to areal densities compatible with single photon emission in the telecom O-band. The PL spectra are identical in ensembles of centers and in single G centers, demonstrating the establishment of a reproducible procedure for the creation of single centers in silicon for quantum technologies.  

\section*{Acknowledgments}
This work was supported by the French National Research Agency (ANR) through the projects ULYSSES (No. ANR-15-CE24-0027-01), OCTOPUS (No. ANR-18- CE47-0013-01) and QUASSIC (ANR-18-ERC2-0005-01), the Occitanie region through the SITEQ contract, the German Research Foundation (DFG) through the ULYSSES project (PE 2508/1-1) and the European Union's Horizon 2020 program through the FET-OPEN project NARCISO (No. 828890) and the \mbox{ASTERIQS} project (Grant No.~820394). The authors thank the Nanotecmat platform of the IM2NP institute, and Louis Hutin and Beno\^{i}t Bertrand (CEA Leti) for their contributions to the preparation of the $^{28}$SOI substrates. A. Durand acknowledges support from the French DGA.

\bibliography{G_implant_biblio}

\end{document}